\documentclass[conference, letterpaper]{IEEEtran}
\IEEEoverridecommandlockouts

\usepackage{cite}
\usepackage{amsmath,amssymb,amsfonts}
\usepackage{graphicx}
\usepackage{textcomp}
\usepackage{xcolor}
\def\BibTeX{{\rm B\kern-.05em{\sc i\kern-.025em b}\kern-.08em
    T\kern-.1667em\lower.7ex\hbox{E}\kern-.125emX}}

\usepackage{kotex}
\usepackage[T1]{fontenc}
\usepackage[utf8]{inputenc} 
\usepackage{xcolor}
\usepackage{algorithm}
\usepackage{algpseudocode}
\usepackage{booktabs}
\usepackage{enumitem}
\usepackage{float}

\newtheorem{remark}{Remark}

\newcommand{\argmin}{{\arg\min}}
\newcommand{\argmax}{{\arg\max}}

\begin{document}

\title{Multipoint Code-Weight Sphere Decoding: Parallel Near-ML Decoding for Short-Blocklength Codes}


\author{\IEEEauthorblockN{Yubeen Jo*, Geon Choi, Yongjune Kim, and Namyoon Lee}
\IEEEauthorblockA{*\textit{Department of Electrical Engineering}, Korea University, Seoul, South Korea \\
\textit{Department of Electrical Engineering}, POSTECH, Pohang, South Korea \\
*jybin00@korea.ac.kr, $\{$geon.choi, yongjune, nylee$\}$@postech.ac.kr}
}

\maketitle

\begin{abstract}
 Ultra-reliable low-latency communications (URLLC) operate with short packets, where finite-blocklength effects make near-maximum-likelihood (near-ML) decoding desirable but often too costly. This paper proposes a two-stage near-ML decoding framework that applies to any linear block code. In the first stage, we run a low-complexity decoder to produce a candidate codeword and a cyclic redundancy  check. When this stage succeeds, we terminate immediately. When it fails, we invoke a second-stage decoder, termed multipoint code-weight sphere decoding (MP-WSD). The central idea behind {MP-WSD} is to concentrate the ML search where it matters. We pre-compute a set of low-weight codewords and use them to generate structured local perturbations of the current estimate. Starting from the first-stage output, MP-WSD iteratively explores a small Euclidean sphere of candidate codewords formed by adding selected low-weight codewords, tightening the search region as better candidates are found. This design keeps the average complexity low: at high signal-to-noise ratio, the first stage succeeds with high probability and the second stage is rarely activated; when it is activated, the search remains localized. Simulation results show that the proposed decoder attains near-ML performance for short-blocklength, low-rate codes while maintaining low decoding latency.
\end{abstract}


\section{Introduction}

  Ultra-reliable low-latency communications (URLLC) operate with short packets, where finite-blocklength effects make strong codes at low rates essential and put the burden squarely on the decoder: it must be both reliable and fast \cite{shtblk_len_ch_code, effi_dec_sht_blk_len_6G, 6G, boss_code,boss2,deep_polar_code,sparse_pretran_polar}. A large body of work has therefore developed suboptimal decoding algorithms that trade off complexity and performance, including sphere decoding (SD) \cite{sphere, SD_CA}, ordered-statistics decoding (OSD) \cite{osd}, guessing random additive noise decoding (GRAND) \cite{orbgrand, product-like, segment_grand}, and list decoding methods such as successive-cancellation list (SCL) decoding \cite{scl}, as well as linearity-enhanced serial list decoders inspired by sphere-based ideas \cite{linearity-enhanced}. Despite this progress, a persistent gap remains: for short blocklengths and low-rate codes, obtaining near-maximum-likelihood (near-ML) performance at a complexity level compatible with URLLC is still challenging.
  
  Recently, code-weight sphere decoding (WSD) was proposed in \cite{wsd} as a universal near-ML decoding method for arbitrary linear block codes. WSD improves a codeword estimate by repeatedly applying local moves that greedily reduce the Euclidean distance (ED) to the channel observation. The difficulty is that the ED landscape over the discrete code lattice is highly non-convex. As a result, a greedy descent can stall at a \emph{local} optimum, missing the ML solution. One can reduce this risk by enlarging the search sphere, but doing so increases memory and computation rapidly (and in practice, prohibitively), which conflicts with tight latency constraints.

In this paper, we address this bottleneck by replacing a single deep greedy search with many shallow searches executed in parallel. We propose \emph{multipoint code weight sphere decoding (MP-WSD)}, a parallel near-ML decoding framework designed for modern receivers with substantial parallel compute resources (e.g., advanced baseband processing in AI-RAN). The decoder begins with a low-complexity list decoder that produces a small set of candidate codewords. These candidates act as \emph{anchors}---multiple starting points from which MP-WSD explores the code lattice in parallel.

The key idea is to exploit the geometric uniformity of linear codes. Around each anchor we form a local Hamming sphere of candidate moves. By linearity, the neighborhood of any codeword is simply a translation of the neighborhood of the all-zero codeword. Consequently, we can pre-compute a compact set of low-weight ``move patterns'' once, and reuse them for every anchor. Each parallel trajectory then performs an iterative hopping procedure: apply a small set of translated patterns, keep the move that most reduces ED, and repeat for a fixed number of iterations.

This multipoint architecture provides three advantages:
\begin{itemize}
    \item \textbf{Escaping local optima via diversity:} multiple anchors increase the chance that at least one trajectory enters the basin of attraction of the ML codeword, without requiring a larger sphere.
    \item \textbf{Predictable latency:} rather than growing the sphere (and the worst-case complexity), MP-WSD fixes the depth of each local search and spreads the work across parallel paths, yielding a deterministic latency budget suitable for URLLC.
    \item \textbf{Embarrassingly parallel implementation:} the searches are independent across anchors, enabling straightforward hardware parallelization with minimal synchronization.
\end{itemize}

Simulation results show that MP-WSD approaches ML performance for a range of low-rate codes in the short-blocklength regime, while significantly reducing the \emph{average} decoding complexity compared with high-order OSD and large-list SCL decoders.

\section{Preliminaries}

    In this section, we establish the notation and fundamental concepts employed throughout this paper.

\subsection{Notations}
    Let $|\mathcal{A}|$ be the cardinality of a set $\mathcal{A}$, and $\|\mathbf{v}\|$ be the $L_2$-norm of a vector $\mathbf{v}$. The support $\text{supp}(\mathbf{c})$ is the set of indices of non-zero elements in $\mathbf{c}$. $\mathbb{F}_2$ denotes the binary Galois field, $\mathbf{1}_N$ is the all-one vector of length $N$, and $\oplus$ represents modulo-2 addition. $d_H(\cdot,\cdot)$ and $w_H(\cdot)$ denote the Hamming distance and weigh, respectively.

\subsection{Channel Coding System}
    Consider a binary linear block code $\mathcal{C}$ with parameters $(N, K)$, generated by a generator matrix $\mathbf{G} \in \mathbb{F}_2^{K \times N}$. An input message $\mathbf{m} \in \mathbb{F}_2^{K}$ is encoded into a codeword $\mathbf{c} \in \mathbb{F}_2^N$ via the linear transformation $\mathbf{c} = \mathbf{m}\mathbf{G}$.
    
    To enhance error detection, a cyclic redundancy check (CRC) can be optionally integrated. In this scheme, the message $\mathbf{m}$ is first precoded into a vector $\mathbf{v} \in \mathbb{F}_2^{K'}$ where $K' = K + K_{\sf crc}$, with $K_{\sf crc}$ denoting the number of parity bits. This precoded vector then serves as the input to the channel encoder.
    
    For transmission, we employ binary phase shift keying (BPSK). We denote the modulated symbol vector corresponding to a binary vector $\mathbf{c}$ as $\mathbf{x}(\mathbf{c}) = \mathbf{1}_N - 2\mathbf{c}$. The modulated signal is transmitted over an additive white Gaussian noise (AWGN) channel, modeled as: $\mathbf{y} = \mathbf{x}(\mathbf{c}) + \mathbf{n}$, where $\mathbf{n} \sim \mathcal{N}(\mathbf{0}, \sigma^2 \mathbf{I}_N)$ is the noise vector with variance $\sigma^2 = N_0/2$. The decoder aims to recover the transmitted codeword $\hat{\mathbf{c}}$ (or message $\hat{\mathbf{m}}$) from $\mathbf{y}$ with minimum error probability.

\subsection{Code-Weight Sphere and Local Geometry}
    In a linear block code $\mathcal{C}$, the decoding search space can be efficiently structured by decomposing the code lattice into Hamming shells based on the code weights. For any reference codeword ${\bf c} \in \mathcal{C}$, we partition the code space into Hamming shells: 
    \begin{equation}
        \mathcal{C}_{d_\ell} ({\bf c}) \triangleq \{ {\bf c}' \in \mathcal{C}: d_H({\bf c}, {\bf c}') = d_\ell\},
    \end{equation} where $d_\ell$ is the $\ell$-th smallest weight in the code's weight spectrum. Suppose the support of the weight spectrum consists of $L+1$ distinct elements arranged in increasing order. The code-weight sphere centered at ${\bf c}$ with a radius index $r$, denoted as $\mathcal{S}_r({\bf c})$, is defined as:
    \begin{align}
        \mathcal{S}_r({\bf c}) = \bigcup_{\ell=0}^{r} \mathcal{C}_{d_\ell}({\bf c}).
    \end{align}
    
    A fundamental property of linear codes is their geometric uniformity; the local neighborhood structure around any codeword is identical to that around the zero codeword $\mathbf{0}$. Specifically, the Hamming shell centered at an arbitrary codeword $\mathbf{c}$ is a coset of the shell centered at $\mathbf{0}$. Since $d_H(\mathbf{c}, \mathbf{c}') = w_H(\mathbf{c} \oplus \mathbf{c}')$, the following affine relationship holds:
    \begin{align}
        \mathcal{S}_r(\mathbf{c}) = \mathbf{c} \oplus \mathcal{S}_r(\mathbf{0}) = \{ \mathbf{c} \oplus \mathbf{c}' : \mathbf{c}' \in \mathcal{S}_r(\mathbf{0}) \}.
    \end{align}
    This property implies that the set of codeword required to traverse the local sphere is invariant to the location of the codeword. Consequently, we can pre-compute and store the low-weight codewords in $\mathcal{S}_r(\mathbf{0})$ offline and apply them to \textit{any} initial estimate $\hat{\mathbf{c}}^{(0)}$ during the online search, enabling efficient parallel searches without dynamic tree construction.

    \begin{figure}[t]
      \centering
      \includegraphics[width=\columnwidth]{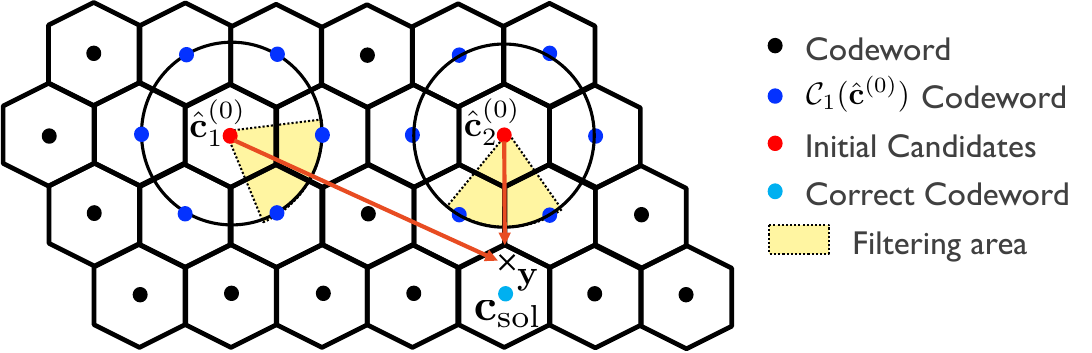}
      \caption{Schematic of the proposed Parallelized WSD. The search starts from multiple initial candidates (red dots) derived from the list decoder, and each path independently converges to a local optimum using the pre-computed code-weight sphere (yellow area).}
      \label{fig:mp_wsd} 
      \vspace{-3mm}
    \end{figure}

\section{Proposed Parallelized Near-ML Decoder}
\label{sec:proposed}

    In this section, we present a two-stage decoding framework that leverages a low-complexity list decoder followed by the proposed Parallel WSD.

\subsection{Initial List Decoding Phase}
    In the initial phase, a low-complexity list decoder, such as SCL or OSD, generates a list of preliminary codeword estimates, denoted as $\mathcal{L}_{\text{init}} = \{ \hat{\mathbf{c}}_1, \hat{\mathbf{c}}_2, \dots, \hat{\mathbf{c}}_L \}$.
    If a CRC is employed, the decoder checks the validity of the most reliable candidate in the list. The process terminates early if the CRC validation passes, significantly reducing the average computational complexity.
    
    In the event of a CRC failure or when a CRC is not available, the system proceeds to the WSD phase using the multiple candidates selected from $\mathcal{L}_{\text{init}}$ as starting points. This list-based initialization is crucial as it prevents the search from being trapped in a local optimum associated with a single erroneous estimate.

\subsection{Multipoint Code-Weight Sphere Decoding Phase}
    The WSD phase encompasses parallel iterative searches designed to refine the set of initial codeword estimates. As illustrated in Fig. \ref{fig:mp_wsd}, the proposed multipoint strategy initiates independent searches from multiple candidates in $\mathcal{L}_{\text{init}}$.
    
    For each path $k \in \{1, \dots, L\}$, the decoder iteratively searches for a better candidate within its local code-weight sphere $\mathcal{S}_{r}(\hat{\mathbf{c}}_k^{(i-1)})$ at the $i$-th iteration.
    The candidate update rule for the $k$-th path is formulated as finding the local minimizer of the Euclidean distance to the received signal $\mathbf{y}$:
    \begin{align}
        \hat{\mathbf{c}}_k^{(i)} = \underset{{{\mathbf{c} \in \mathcal{S}_{r}(\hat{\mathbf{c}}_k^{(i-1)})}}}{\argmin} \|\mathbf{y} - \mathbf{x}(\mathbf{c})\|.
        \label{eq:wsd_optimization}
    \end{align}
    This optimization is performed efficiently using the pre-computed patterns from $\mathcal{S}_r(\mathbf{0})$ and the correlation-based filtering described in Section \ref{subsec:filtering}.

    If the metric improves (i.e., $\|\mathbf{y} - \mathbf{x}(\hat{\mathbf{c}}_k^{(i)})\| < \|\mathbf{y} - \mathbf{x}(\hat{\mathbf{c}}_k^{(i-1)})\|$), the search center is updated, and the process continues to the next iteration. Since all $L$ paths operate independently, they can be executed in parallel on hardware. Finally, the decoder selects the candidate with the minimum Euclidean distance among all converged paths as the final decision.

    \begin{algorithm}[!t]
        \small 
        \caption{\strut{Multipoint Code-Weight Sphere Decoding}}\label{alg:pm_wsd}
        \begin{algorithmic}[1]
        \Require Received signal $\mathbf{y}$, Sphere $\mathcal{S}_r(\mathbf{0})$, Initial list $\mathcal{L}_{\text{init}}$, Max iter $J$, Filter $m$.
        \Ensure Final estimate $\hat{\mathbf{c}}_{\text{best}}$.
        
        \State \textbf{Init:} $\hat{\mathbf{c}}_k^{(0)} \leftarrow \mathcal{L}_{\text{init}}\{k\}$, $\text{active}_k \leftarrow \text{true}$ for all $k=1 \dots |\mathcal{L}_{\text{init}}|$.
        
        \For{$i = 1$ to $J$}
            \State $\text{updated} \leftarrow \text{false}$
            \For{$k = 1$ to $|\mathcal{L}_{\text{init}}|$} \textbf{in parallel}
                \If{$\text{not active}_k$} \textbf{continue} \EndIf
                
                \State \textbf{Filter:} $\mathcal{M}_{k} \leftarrow \underset{{\mathcal{A}\subseteq \mathcal{S}_r(\mathbf{0}),|\mathcal{A}|=m}}{\argmax}\;\;\sum_{c\in \mathcal{A}} \mathcal{G}(\mathbf{c})$ 
                \State \textbf{Search:} $\mathbf{c}^* \leftarrow \underset{{\mathbf{c} \in \mathcal{M}_{k}}}{\argmin} \quad\|\mathbf{y} - \mathbf{x}(\hat{\mathbf{c}}_k^{(i-1)} \oplus \mathbf{c})\|$
                
                \If{$\|\mathbf{y} - \mathbf{x}(\hat{\mathbf{c}}_k^{(i-1)} \oplus \mathbf{c}^*)\| < \|\mathbf{y} - \mathbf{x}(\hat{\mathbf{c}}_k^{(i-1)})\|$}
                    \State $\hat{\mathbf{c}}_k^{(i)} \leftarrow \hat{\mathbf{c}}_k^{(i-1)} \oplus \mathbf{c}^*$, $\text{updated} \leftarrow \text{True}$
                \Else
                    \State $\hat{\mathbf{c}}_k^{(i)} \leftarrow \hat{\mathbf{c}}_k^{(i-1)}$, $\text{active}_k \leftarrow \text{False}$
                \EndIf
            \EndFor
            \If{not $\text{updated}$} \textbf{break} \EndIf
        \EndFor
        \State \Return $\hat{\mathbf{c}}_{\text{best}} \leftarrow \argmin_{\hat{\mathbf{c}}_k^{(i)}} \|\mathbf{y} - \mathbf{x}(\hat{\mathbf{c}}_k^{(i)})\|$
        \end{algorithmic}
    \end{algorithm}


\begin{remark}[Decoding Stability and Refinement]
    By construction, each of the $L$ parallel decoding trajectories ensures a strictly monotonic improvement in reliability (i.e., $\|\mathbf{y} - \mathbf{x}(\hat{\mathbf{c}}_k^{(0)})\| > \dots > \|\mathbf{y} - \mathbf{x}(\hat{\mathbf{c}}_k^{(J)})\|$). 
    This independent greedy update prevents error accumulation within each path. Consequently, the decoder converges to a set of local optima, from which the global best is selected, guaranteeing that the final performance remains at least equal to the best candidate in the initial list.
\end{remark}

\begin{remark}[Difference from Sphere Decoding]
    While both SD and WSD pursue the ML solution, they differ fundamentally. First, unlike SD which typically centers the search on the continuous signal $\mathbf{y}$ (requiring online tree construction), multipoint WSD initiates searches from multiple discrete codeword estimates. This allows WSD to exploit code linearity and construct the search spaces instantly using \textit{offline pre-computed} spheres. Second, instead of a static tree traversal, WSD employs parallel \textit{iterative hopping} strategies (analogous to discrete gradient descent) starting from distinct anchors. Finally, WSD enforces a strict latency bound via the iteration parameter $J$, whereas SD's complexity varies with channel noise.
\end{remark}

\subsection{Complexity Analysis} \label{subsec:filtering}
    To ensure a fair comparison across different decoding architectures, we evaluate the computational complexity in terms of floating-point operations (FLOPs) and normalize them into equivalent ED units.
    We define one ED unit as the cost of computing $\|\mathbf{y} - \mathbf{x}\|^2$ for a vector of length $N$, which requires approximately $3N$ FLOPs.
    Consequently, the complexity of the brute-force ML decoder is normalized to $2^K$ ED units. 

    For the benchmark SCL decoder, we model the complexity of log-likelihood ratio (LLR)-based SCL decoding as $\mathcal{C}_{\sf scl} \approx \frac{4}{3} L \log_2 N$, assuming 4 FLOPs per processing node. 
    This is extended to SCL-back propagation parity check (SCL-BPC) \cite{deep_polar_code} by adding the pre-transformation overhead, modeled as $\mathcal{C}_{\sf bpc} \approx \mathcal{C}_{\sf scl} + \frac{4}{3N} \sum_\ell N_\ell \log_2 N_\ell$. 
    The OSD complexity is approximated by the order-$k$ reprocessing stage, $\mathcal{C}_{\sf osd} \approx \sum_{i=0}^k \binom{K}{i}$. Detailed formulations are summarized in Table \ref{tab:complexity}.

    To reduce the search overhead, we adopt the \textit{correlation-based filtering} technique introduced in \cite{wsd}. As shown in \cite{wsd}, minimizing the Euclidean distance is equivalent to maximizing the correlation for BPSK signals.
    
    In our multipoint framework, we calculate the gain metric $\mathcal{G}_k(\mathbf{c})$ for the $k$-th path ($k=1,\dots,|\mathcal{L}_{\rm init}|$), which represents the correlation improvement by shifting the current center $\hat{\mathbf{x}}_k^{(i-1)}$ by a low-weight codeword $\mathbf{c}$:
    \begin{align}
        \mathcal{G}_k(\mathbf{c}) = \sum_{j \in \text{supp}(\mathbf{c})} -2 y_j \hat{x}_{k,j}^{(i-1)}.
    \end{align}
    Using this metric, we efficiently identify the subset $\mathcal{M}_k \subset \mathcal{S}_r(\mathbf{0})$ containing the top-$m$ candidates without full distance calculations. This operation requires only $w_H(\mathbf{c})$ additions, significantly reducing the computational load.
    
    Considering the filtering and sorting overheads per single search path, the worst-case normalized complexity of a WSD trajectory is bounded by:
    \begin{align}
        \mathcal{C}_{\sf wsd} \approx J \times \left( \underbrace{m\left(1+\frac{1}{3N}\right)}_{\text{\scriptsize Exact ED \& Min}} + \underbrace{\frac{|\mathcal{S}_r(\mathbf{0})|(\bar{w} + \log_2 m)}{3N}}_{\text{\scriptsize Filtering \& Sorting}} \right),
    \end{align}
    where $\bar{w}$ denotes the average Hamming weight of the candidate codewords in $\mathcal{S}_r(\mathbf{0})$ and $m$ is the filtering threshold size ($m=|\mathcal{M}_k|$).
    Finally, the total worst-case complexity is given by integrating the parallel branches:
    \begin{align}
        \mathcal{C}_{\sf mp-wsd} = \mathcal{C}_{\sf init} + P_{\sf act} \times |\mathcal{L}_{\text{init}}| \times \mathcal{C}_{\sf wsd},
    \end{align}
    where $P_{\sf act}$ denotes the activation probability of the WSD phase. For CRC-aided systems, $P_{\sf act} = P_{\sf e,crc}$ represents the probability of a CRC failure. For scenarios without CRC (e.g., the Reed-Muller (RM) code with OSD), $P_{\sf act}$ is effectively set to $1$, as the WSD stage can be configured to operate in an \textit{always-on mode} (AOM) to guarantee reliability refinement.

    At high SNR, $P_{\sf e,crc} \approx 0$, making the overhead negligible. At low SNR, practical feasibility is ensured via three mechanisms:
    \begin{itemize}
        \item \textbf{Bounded Latency}: The worst-case delay is deterministically limited by $J$.
        \item \textbf{Hierarchical Parallelizability}: The architecture supports parallelism at two levels: (1) \textit{Inter-path parallelism}, where $|\mathcal{L}_{\text{init}}|$ decoding trajectories operate independently, and (2) \textit{Intra-path parallelism}, where filtering and ED calculations for candidates within $\mathcal{M}_k$ are performed concurrently.
        \item \textbf{Adaptivity}: WSD can be selectively disabled if the estimated SNR falls below a threshold to prevent unnecessary overhead.
    \end{itemize}

\begin{table}[htbp]
    \renewcommand{\arraystretch}{1.3}
    \centering
    \caption{Theoretical Normalized Complexity Analysis (in ED units)}
    \label{tab:complexity}
    \footnotesize
    \begin{tabular}{@{}ll@{}}
        \toprule
        Decoder & \multicolumn{1}{c}{Normalized Complexity (Approx.)} \\ 
        \midrule
        SCL$(L)$            & $\mathcal{C}_{\sf scl} \approx \frac{4}{3} L \log_2 N$  \\ 
        SCL$(L)$ + MP-WSD$(r, J, L_{\text{init}})$ & $\mathcal{C}_{\sf scl} + P_{\sf act} \times L_{\text{init}} \times \mathcal{C}_{\sf wsd}$ \\ 
        \midrule
        SCL-BPC$(L)$        & $\mathcal{C}_{\sf bpc} \approx \mathcal{C}_{\sf scl} + \frac{4}{3N} \sum_{\ell} N_\ell \log_2 N_\ell$ \\
        SCL-BPC$(L)$ + MP-WSD$(r, J, L_{\text{init}})$ & $\mathcal{C}_{\sf bpc} + P_{\sf act} \times L_{\text{init}} \times \mathcal{C}_{\sf wsd}$ \\
        \midrule
        OSD$(k)$            & $\mathcal{C}_{\sf osd} \approx \sum_{i=0}^k \binom{K}{i}$ \\
        OSD$(k)$ + MP-WSD$(r, J, L_{\text{init}})$ & $\mathcal{C}_{\sf osd} + L_{\text{init}} \times \mathcal{C}_{\sf wsd}$ \\
        \midrule
        MLD & $2^K$ \\ 
        \bottomrule
    \end{tabular}
    
    \vspace{1mm} 
    \par 
    \scriptsize 
    \raggedright 
    * $L_{\text{init}} = |\mathcal{L}_{\text{init}}|$: Number of parallel paths derived from the initial stage. \\
    * $\mathcal{C}_{\mathsf{wsd}}$: Complexity per single WSD trajectory (Eq. 6).
\end{table}

\section{Simulation Results}\label{sec:simulation}

In this section, we evaluate the block error rate (BLER) and computational complexity of the proposed MP-WSD framework. To demonstrate the universality and robustness of our approach, we consider three distinct classes of linear block codes: CRC-aided (CA) polar codes, CA-deep polar (CA-DP) codes, and RM codes.

\subsection{Simulation Settings and Benchmarks}
The simulation environments and parameters are configured as follows:

\begin{itemize}
    \item \textbf{CA-polar code:} We construct (256, 16) CA-polar code with 5G-NR reliability sequence. An 11-bit CRC with generator polynomial $g(x) = 1+x^5+x^9+x^{10}+x^{11}$ is employed for early termination and candidate selection \cite{3gpp2020}.
    
    \item \textbf{CA-DP code:} For CA-DP codes \cite{deep_polar_code}, which involve pre-transformation to enhance the minimum distance, we utilize the SCL-BPC decoder as the initial stage. The simulations focus on the $(128, 16)$ configuration.
    
    \item \textbf{RM code:} To verify the applicability of MP-WSD to non-polar structures, we evaluate the $\mathcal{RM}(128, 29)$ code. The OSD \cite{osd} with orders $k \in \{2, 3, 4\}$ serves as the initial decoding stage.
    
    \item \textbf{Proposed MP-WSD:} The maximum number of iterations is set to $J=4$, based on empirical observations that performance gains saturate beyond this point. We analyze the performance-complexity tradeoff by varying the search radius $r$ and the number of multipoint initialization paths $|\mathcal{L}_{\text{init}}|$. The initialization list $\mathcal{L}_{\text{init}}$ is constructed by selecting the most reliable candidates from the initial decoder. Specifically, for SCL (upon CRC failure), candidates are selected in ascending order of their path metrics, while for OSD, candidates are chosen based on the smallest Euclidean distance among the generated patterns. regarding the filtering parameter, we set $m=\max(10,\lceil0.02\times|\mathcal{S}_r(\mathbf{0})|\rceil)$ to prune the search space efficiently without performance degradation.
    
    \item \textbf{Reference:} The maximum likelihood decoder (MLD) performance and theoretical bounds—specifically the random coding union (RCU) bound \cite{rcu} and the meta-converse bound \cite{meta_conv}—are provided as benchmarks for optimality.
\end{itemize}

Table \ref{tab:simulation_parameters} details the cardinality of the code-weight spheres used in the simulations.

    \begin{table}[tbph]
    \centering
    \caption{Cardinality of the code-weight sphere $\mathcal{S}_r({\mathbf{0}})$ in Sec.~\ref{sec:simulation}}
    \label{tab:simulation_parameters}
    \resizebox{0.95\columnwidth}{!}{%
        \begin{tabular}{@{}lllcccc@{}}
    \toprule
    & $(N,K)$ & Fig.&  $r=1$ & $r=2$ & $r=3$ & $r=4$ \\ 
    \midrule
    CA-polar &$(256,16)$ &\ref{fig:sim_scl_wsd} & 2 & 11 & \textbf{538} & \textbf{6,472}\\ 
    CA-DP  &$(128,16)$  & \ref{fig:sim_deep_polar_wsd}& \textbf{611} & \textbf{14,491} & 51,060 & $-$\\ 
    $\mathcal{RM}$ & $(128,29)$ & \ref{fig:sim_osd_rm_wsd} & \textbf{10,688}& $-$&$-$ &$-$ \\
    \bottomrule
    \end{tabular}
    }
    \vspace{-3mm}
    \end{table}

    \begin{figure}[htbp]
      \centering
      \includegraphics[width=\columnwidth]{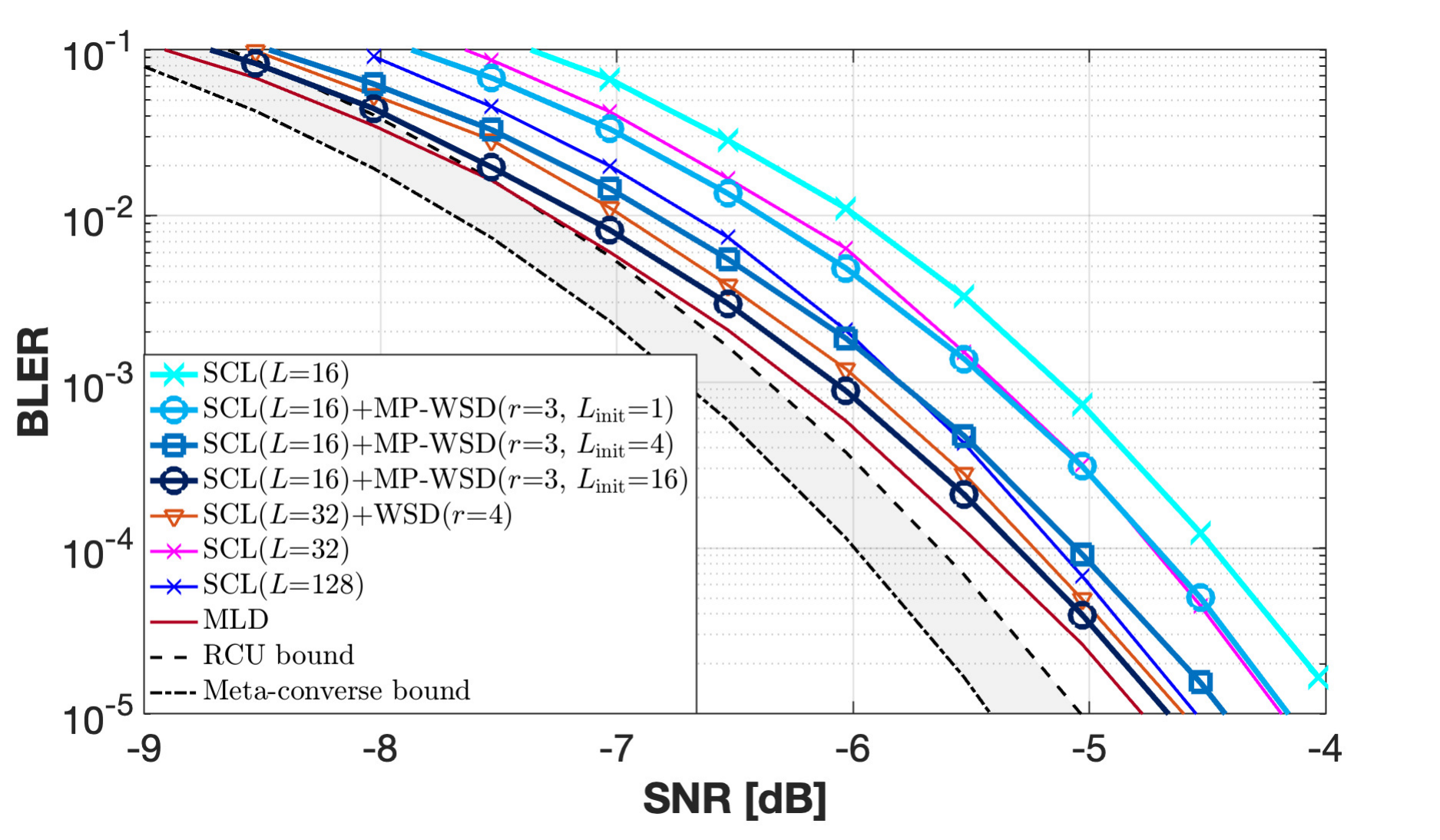}
      \vspace{-6mm}
      \caption{BLER performance of CA-polar codes ($N=256, K=16$) with SCL and MP-WSD. Notably, SCL($L=16$) combined with MP-WSD achieves performance comparable to the higher-complexity SCL($L=32$).}
      \label{fig:sim_scl_wsd}
      \vspace{-3mm}
    \end{figure}

    \begin{figure}[htbp]
      \centering
      \includegraphics[width=\columnwidth]{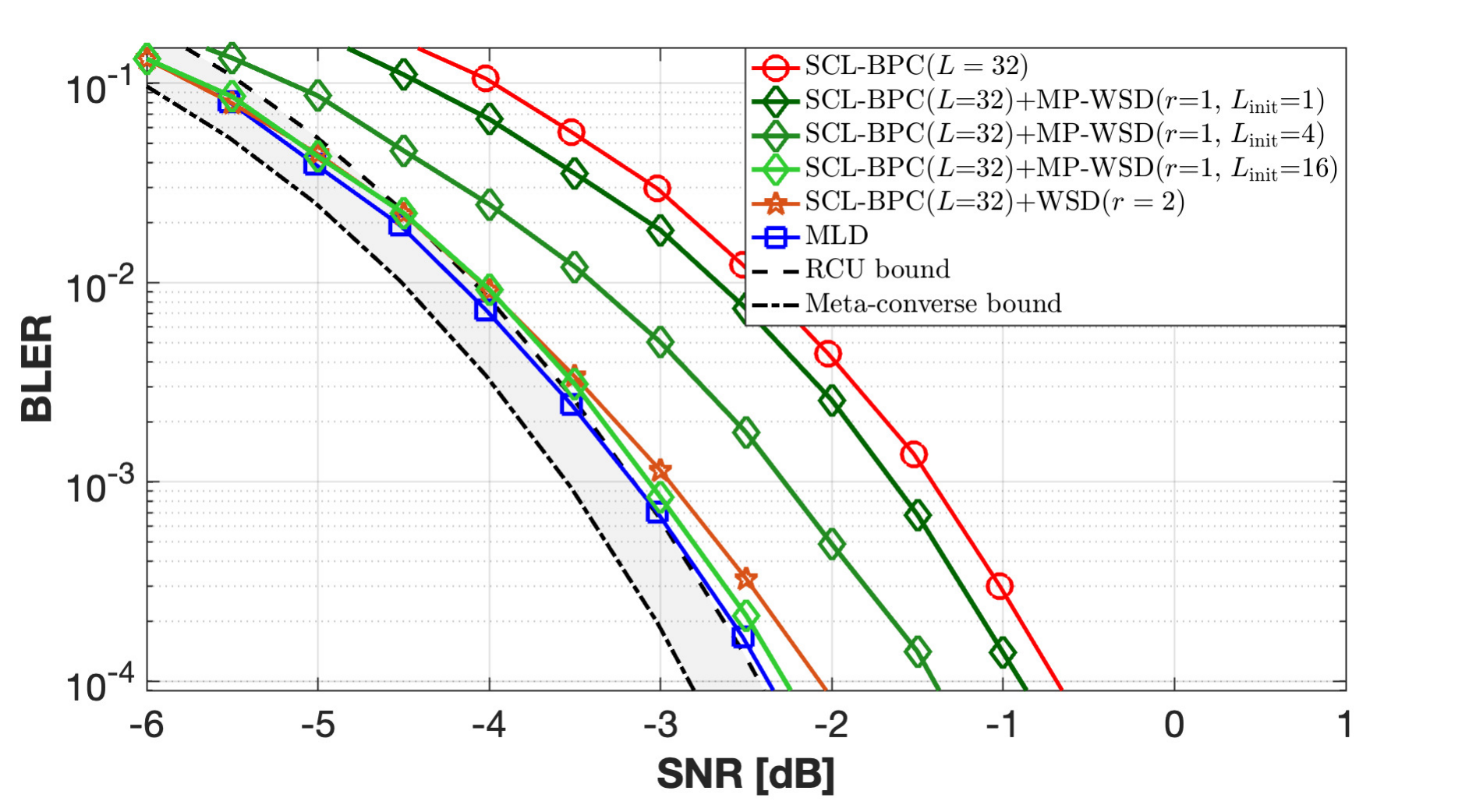}
      \vspace{-6mm}
      \caption{BLER performance of CA-DP codes ($N=128, K=16$) with varying number of initialization paths $|\mathcal{L}_{\text{init}}|$. MP-WSD with $|\mathcal{L}_{\text{init}}|=16$ significantly outperforms the baseline SCL-BPC($L=32$).}
      \label{fig:sim_deep_polar_wsd}
      \vspace{-3mm}
    \end{figure}

    \begin{figure}[htbp]
      \centering
      \includegraphics[width=\columnwidth]{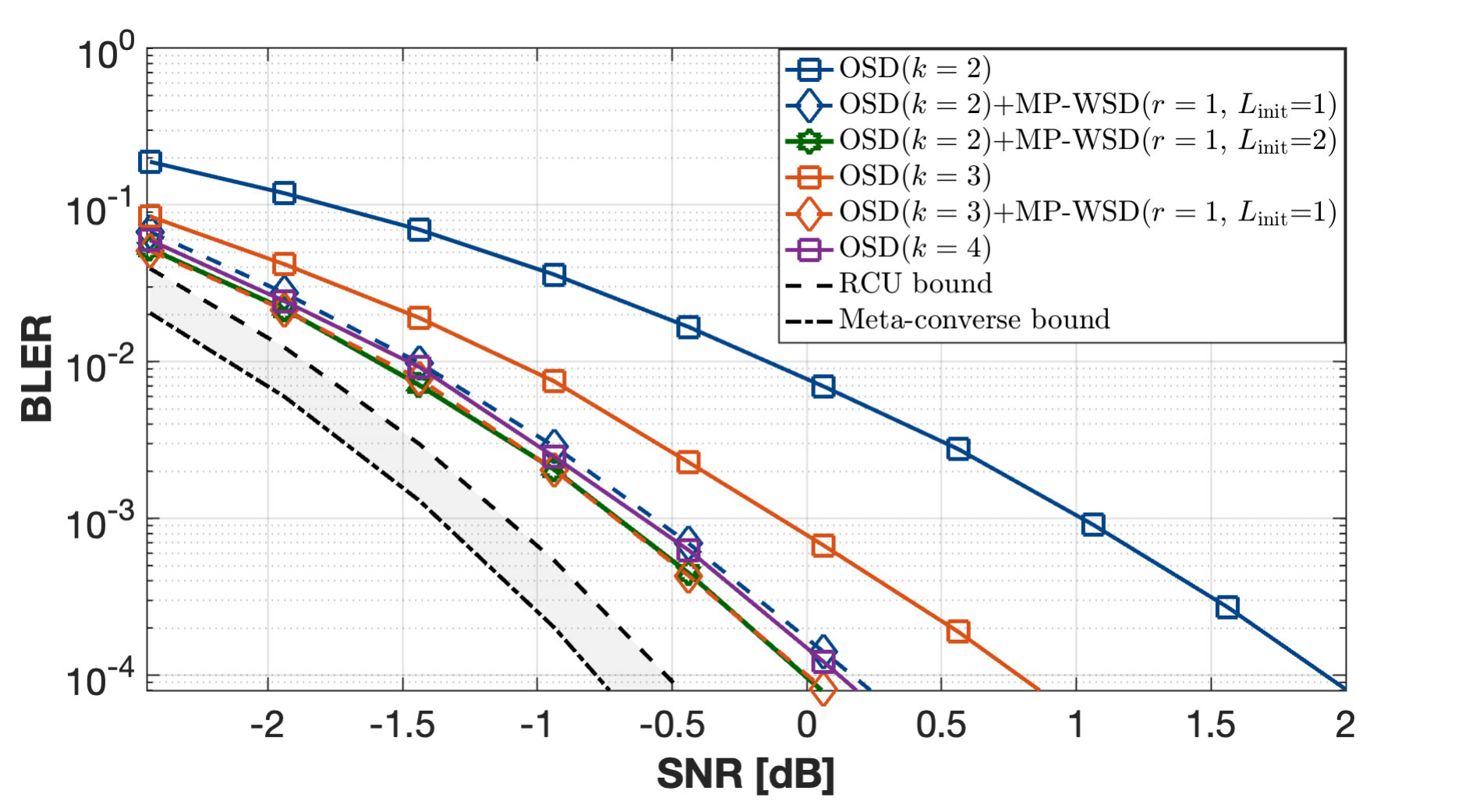}
      \vspace{-6mm}
      \caption{BLER performance of $\mathcal{RM}(128, 29)$ codes. MP-WSD with a low-order OSD($k=2$) initialization approaches the reliability of the high-complexity OSD($k=4$).}
      \label{fig:sim_osd_rm_wsd}
      \vspace{-3mm}
    \end{figure}

\subsection{BLER Performance and Diversity Gain}

Fig. \ref{fig:sim_scl_wsd} presents the BLER performance for CA-polar codes. A critical observation is the efficacy of the multipoint strategy. The proposed SCL($L=16$) + MP-WSD scheme achieves a BLER performance nearly identical to that of the baseline SCL($L=32$), and effectively closes the gap to the MLD bound as the search radius $r$ increases. This implies that a lower-complexity initial decoder (smaller list size), when augmented with localized sphere searches, can match the reliability of significantly more complex serial decoders. The multiple starting points ($\mathcal{L}_{\text{init}}$) allow the decoder to explore distinct neighborhoods of the code lattice, thereby increasing the probability of converging to the global ML solution even if the primary candidate is suboptimal.

For the $(128, 16)$ CA-DP code shown in Fig. \ref{fig:sim_deep_polar_wsd}, we explicitly compare the effect of the number of initialization paths, $|\mathcal{L}_{\text{init}}| \in \{1, 4, 16\}$. It is observed that increasing $|\mathcal{L}_{\text{init}}|$ with a shallow search radius ($r=1$) yields performance comparable to that of a single-point search with a larger radius. Specifically, MP-WSD with $|\mathcal{L}_{\text{init}}|=16$ outperforms the standard SCL-BPC($L=32$) by approximately 1.2 dB at a BLER of $10^{-3}$, essentially achieving MLD performance. 

    \begin{figure}[htbp]
      \centering
      \includegraphics[width=\columnwidth]{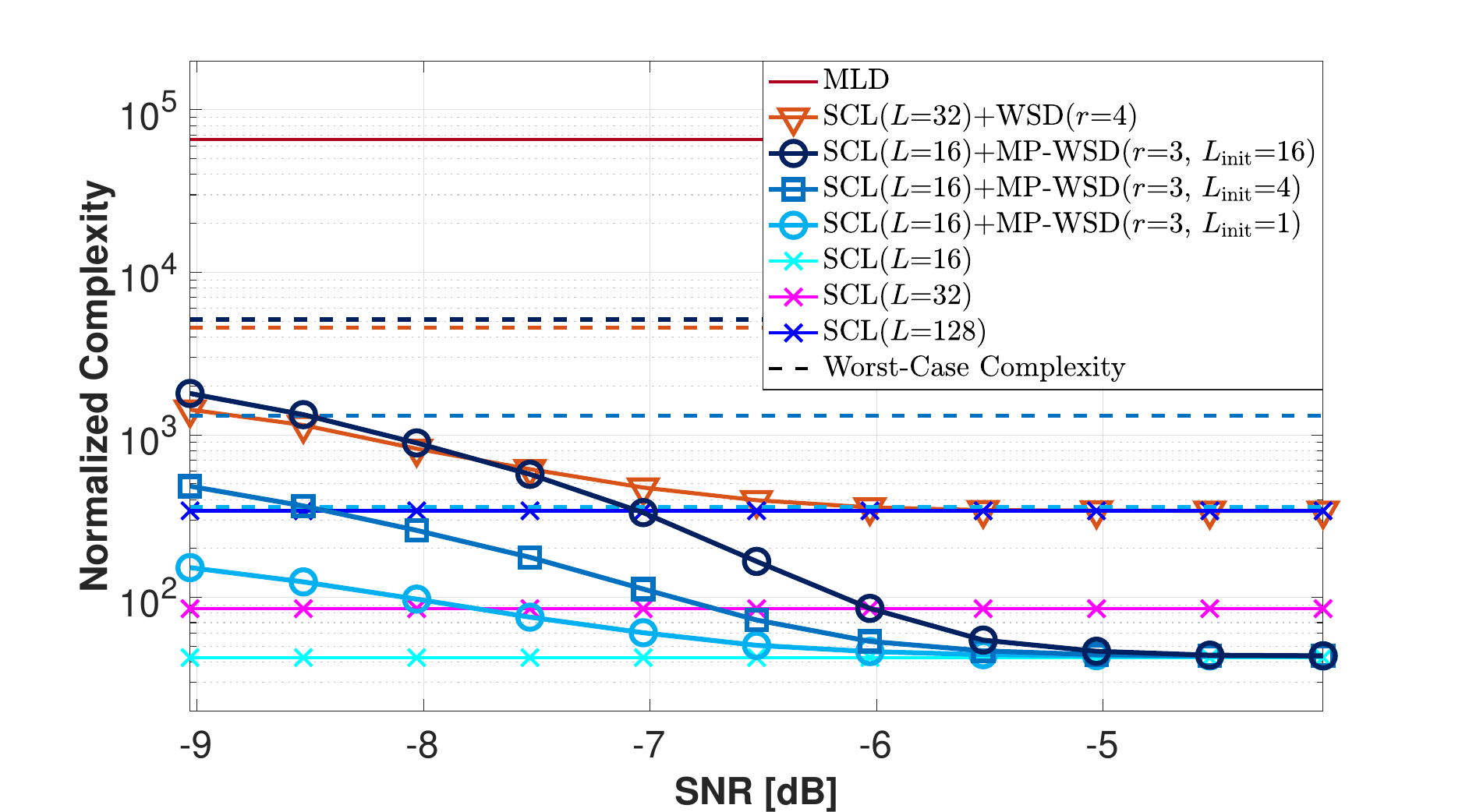}
      \vspace{-3mm}
      \caption{Average normalized complexity for CA-polar codes ($N=256, K=16$). The total complexity adaptively converges to the baseline SCL complexity as SNR increases.}
      \label{fig:sim_scl_wsd_comp}
      \vspace{-3mm}
    \end{figure}

    \begin{figure}[htbp]
      \centering
      \includegraphics[width=\columnwidth]{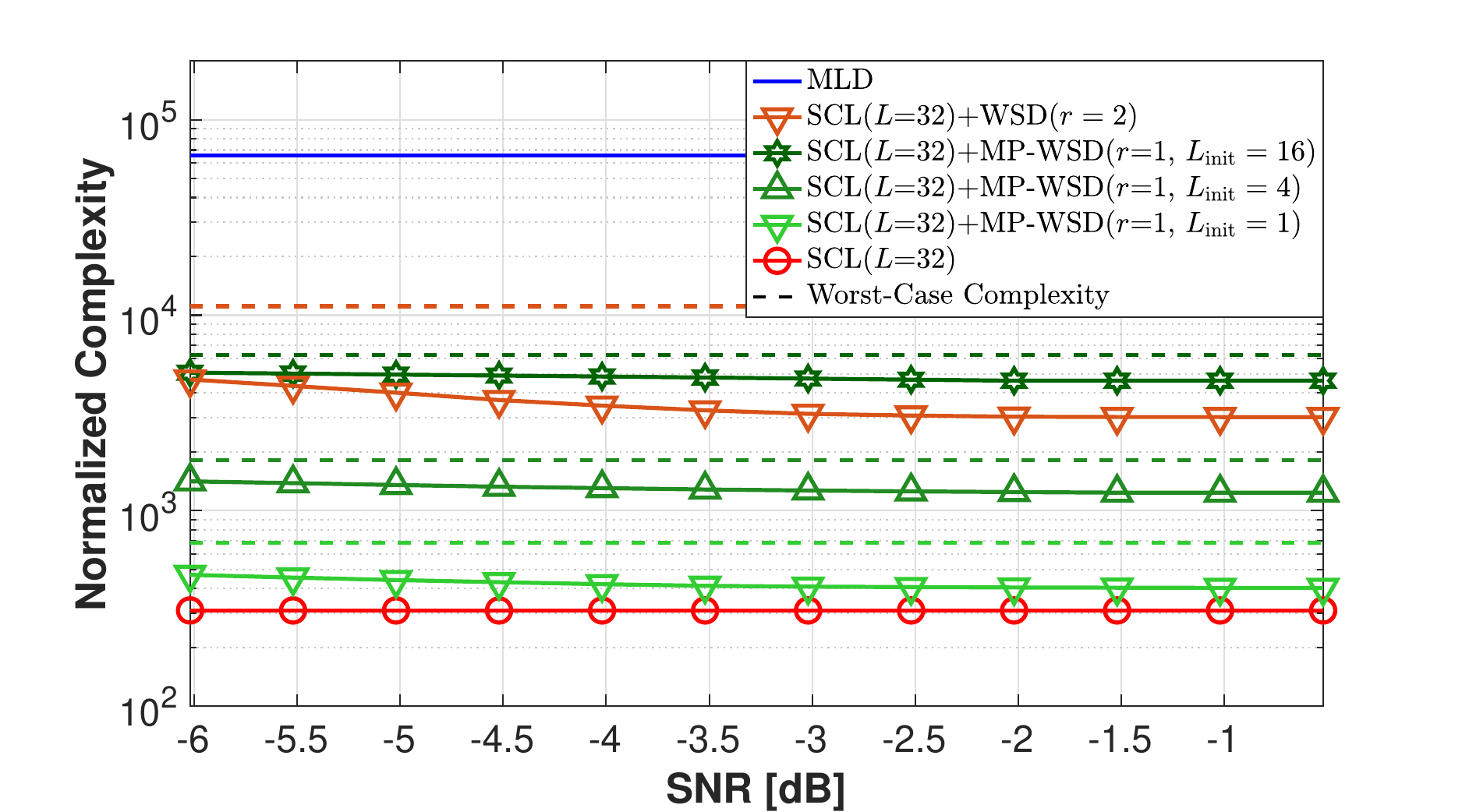}
      \vspace{-3mm}
      \caption{Average normalized complexity for CA-DP codes ($N=128, K=16$). While MP-WSD utilizes multiple paths, the average computational load remains low in the operational SNR regime.}
      \label{fig:sim_scl_bpc_wsd_comp}
      \vspace{-3mm}
    \end{figure}

    \begin{figure}[htbp]
      \centering
      \includegraphics[width=\columnwidth]{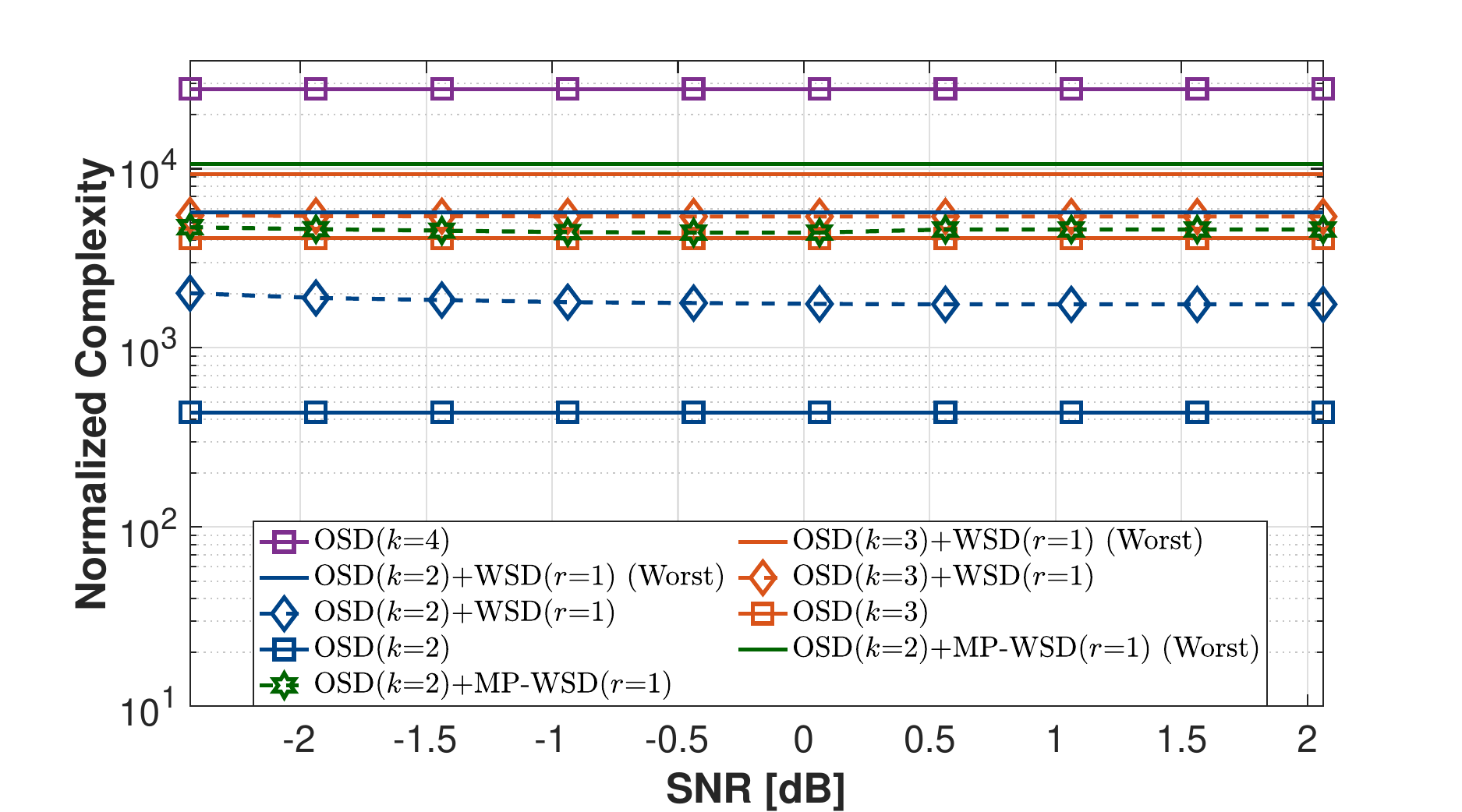}
      \caption{Average normalized complexity for $\mathcal{RM}(128, 29)$ codes. MP-WSD demonstrates orders of magnitude lower complexity compared to high-order OSD ($k=4$).}
      \label{fig:sim_osd_rm_wsd_comp}
      \vspace{-3mm}
    \end{figure}
    
\subsection{Complexity Analysis and Latency Trade-off}

    Figs. \ref{fig:sim_scl_wsd_comp} to \ref{fig:sim_osd_rm_wsd_comp} present the normalized computational complexity. 
    A key structural advantage lies in the trade-off between total operations and latency. While \textit{total} computational cost sums over all $|\mathcal{L}_{\text{init}}|$ paths, the \textit{latency} is significantly reduced due to parallelization. Since each path operates independently on a small sphere, the critical path latency is governed by the shallow search complexity rather than the cumulative cost of a large serial list or deep tree search. This makes MP-WSD particularly suitable for URLLC applications where deterministic low latency is mandatory.

\section{Conclusion}
    In this paper, we proposed a parallelized near-ML decoding framework that addresses the stringent reliability and latency requirements of URLLC in the finite blocklength regime. Our approach, MP-WSD, adaptively combines a low-complexity initial list decoder with an iterative refinement strategy starting from multiple anchors. By exploiting the geometric uniformity of linear codes and initiating independent searches in parallel, MP-WSD effectively mitigates the local optimum problem of greedy searches without the exponential complexity of deep tree-based methods. Simulation results across various code structures—including CA-polar, CA-DP, and RM codes—confirm that MP-WSD achieves near-ML performance with a deterministic latency bound. This parallel architecture established in this work provides a promising foundation for implementing high-throughput, near-ML decoders in future reliability-critical communication systems.



\IEEEtriggeratref{10}
\bibliographystyle{IEEEtran}
\bibliography{conf}

\vspace{12pt}

\end{document}